\pdfoutput=1
\documentclass[10pt,a4paper]{article}
\usepackage[T2A]{fontenc}
\usepackage[utf8]{inputenc}
\usepackage[english]{babel}
\usepackage{amssymb,graphicx}
\usepackage{amsmath,amsfonts}
\usepackage{amsthm}
\usepackage{framed}
\usepackage{xcolor}
\usepackage{fullpage}
\usepackage[makeroom]{cancel}
\usepackage[export]{adjustbox} 
\usepackage{microtype}
\usepackage{hyperref}

 \newcommand{\Tr}{\mathop{\mathrm{Tr}}\nolimits}

 \newtheorem{thm}{Theorem}

\title{\textbf{Spiralling branes and $R$-matrices
  }}
\author{Yegor
  Zenkevich\thanks{yegor.zenkevich@gmail.com}
  \footnote{On leave from ITMP MSU.}\\
  {\small\textit{Departments of Mathematics, University of
      California, Berkeley, USA}}} \date{}
\begin{document}
\maketitle
\vspace{-31ex}
\noindent
{\textit{To T.}}
\vspace{25ex}

\begin{abstract}
  We extend the dictionary between Type IIB branes and representations
  of the Ding-Iohara-Miki (DIM) algebra to the case when one of the
  space directions is a circle. It is well-known that the worldvolume
  theory on branes wrapping the circle is a $5d$ $\mathcal{N}=1$ gauge
  theory with adjoint matter, or more generally of cyclic quiver type,
  and the corresponding intertwiners of the DIM algebra give their
  Nekrasov partition functions. However, we find that there exists a
  much wider natural class of intertwiners corresponding to branes
  spiralling around the compactified direction, with many interesting
  properties. We consider two examples, one corresponding to a
  spiralling D5 brane and another to a D3 brane. The former gives rise
  to the $K$-theoretic vertex function counting sheaves on
  $\mathbb{C}^3$ while the latter produces the ``non-stationary
  elliptic Ruijsenaars wavefunctions'' introduced recently by
  Shiraishi.
\end{abstract}




  



\section{Introduction}
\label{sec:introduction}
String theory branes can be understood on several levels. Historically
they were first introduced as boundary conditions for open strings in
which case they can be thought of as submanifolds of the ambient (or
bulk) space, e.g.~$\mathbb{R}^{10}$. However, this plain geometric
description becomes inadequate if a stack of $N>1$ branes spanning the
same submanifold is considered. Then, somewhat akin to Heisenberg's
development of quantum mechanics from the classical picture, the
coordinates normal to the branes become matrices acting in an
$N$-dimensional vector space (the Chan-Paton
space)~\cite{Polchinski:1998rq}. Reformulated more mathematically, a
stack of branes becomes a rank $N$ vector bundle on a submanifold of
the bulk space. Naturally, the bundle may be topologically nontrivial
and thus some stacks of branes wrapping the same submanifold may be
nonequivalent.

More generally, one would like to consider the possibility of
antibranes, i.e.~the objects which upon being added to the system
``cancel'' (annihilate) branes of a given type. Since there are
nonequivalent stacks of branes that may live on the same submanifolds
one can ask what happens if one tries to annihilate a pair of such
inequivalent stacks. The result turns out to be a stack of branes on a
sub-submanifold, in general of a smaller dimension. The mathematical
formalism to consider branes of different dimension on equal footing
is provided by coherent sheaves on the bulk
space~\cite{Sharpe:2003dr}. Vector bundles on submanifolds of the bulk
space are examples of coherent sheaves, however the sheaf approach is
more general. The essence of it is to understand vector bundles as
modules over the ring of functions on the bulk space. Indeed, if
$v(x)$ is a (local) section of a bundle $\mathcal{E}$ then $f(x) v(x)$
is too for every function $f(x)$ on the bulk space. Adding
non-interacting branes to the system corresponds to taking the direct
sum of sheaves while the annihilation is captured by introducing
equivalence relation which allows one to consider complexes of
sheaves.

The overall picture with modules over a ring, sums and complexes
thereof begs for a representation theoretic description. It does arise
on a more refined level when one considers \emph{bound states} of
branes. The representation space is interpreted as the Hilbert space
of (BPS) bound states of a ``heavy'' brane with a number of ``light''
branes.  The algebras rising in this way are called the BPS
algebras~\cite{Harvey:1996gc}. Pairs of ``heavy'' branes correspond to
tensor products of representations.

The BPS algebras arise in different string theories and in different
backgrounds. Here we will be interested in a particular
case~\cite{Zenkevich:2018fzl} in which branes are of Type IIB string
theory and the algebra is the Ding-Iohara-Miki (DIM)
algebra~\cite{DIM}\footnote{More generally, also of other quantum
  toroidal algebras~\cite{Zenkevich:2019ayk} and affine
  Yangians~\cite{Gaiotto:2017euk},~\cite{Galakhov:2018lta}.}. In this
approach 3- and 5-branes correspond to two different types of DIM
algebra representations, collections of non-interacting branes are
tensor products of representations and brane junctions are
intertwining operators between ingoing and outgoing collections of
branes. Also, brane \emph{crossings} (when one brane passes behind
another without merging) correspond to DIM $R$-matrices taken in the
appropriate representations. The details about the DIM algebra, its
representations and some of the intertwining operators can be found
e.g.\ in the Appendices
to~\cite{Zenkevich:2018fzl},~\cite{Zenkevich:2022dju}
and~\cite{Bayindirli:2023byn}.

To make the correspondence more precise in the table below we list the
main types of branes and representations we are going to consider.
\begin{equation}
  \label{eq:1}
    \begin{array}{l|l|ccc|cc|c|c}
      &&&&&\multicolumn{2}{c|}{\text{picture}}&\tau&\\
      \text{Brane}& \text{DIM rep} &\textcolor{blue}{\mathbb{C}_q}&\textcolor{red}{\mathbb{C}_{t^{-1}}}&\textcolor{violet}{\mathbb{C}_{t/q}}&\mathbb{R}_x&S^1_y&\mathbb{R}_{\tau}&S^1_{10}\\
    \hline
    \textcolor{violet}{\mathrm{D5}_{q,t^{-1}}} &
    \textcolor{violet}{\mathcal{F}^{(0,1)}_{q,t^{-1}}}&--&--&&&-&&-\\
    \textcolor{red}{\mathrm{D5}_{q,t/q}} &
    \textcolor{red}{\mathcal{F}^{(0,1)}_{q,t/q}}&--&&--&&-&&-\\
    \textcolor{blue}{\mathrm{D5}_{t^{-1},t/q}} &
    \textcolor{blue}{\mathcal{F}^{(0,1)}_{t^{-1},t/q}}&&--&--&&-&&-\\
    \hline
    \textcolor{violet}{\mathrm{NS5}_{q,t^{-1}}} &
    \textcolor{violet}{\mathcal{F}^{(1,0)}_{q,t^{-1}}}&--&--&&-&&&-\\
    \textcolor{red}{\mathrm{NS5}_{q,t/q}} &
    \textcolor{red}{\mathcal{F}^{(1,0)}_{q,t/q}}&--&&--&-&&&-\\
    \textcolor{blue}{\mathrm{NS5}_{t^{-1},t/q}} &
    \textcolor{blue}{\mathcal{F}^{(1,0)}_{t^{-1},t/q}}&&--&--&-&&&-\\
    \hline
    \textcolor{violet}{\mathrm{D3}_{t/q}} &
    \textcolor{violet}{\mathcal{V}_{t/q}}&&&--&&&-&-\\
    \textcolor{red}{\mathrm{D3}_{t^{-1}}} &
    \textcolor{red}{\mathcal{V}_{t^{-1}}}&&--&&&&-&-\\
    \textcolor{blue}{\mathrm{D3}_q} &
    \textcolor{blue}{\mathcal{V}_q}&--&&&&&-&-\\
  \end{array} 
\end{equation}
In the setup above we consider branes on $\mathbb{C}^3\times
\mathbb{R}^2 \times S^1 \times S^1$ and we have turned on the
$\Omega$-background along the three $\mathbb{C}$ planes, its
parameters being $q$, $t^{-1}$ and $t/q$ respectively. The ``picture''
label indicates what directions we are going to draw in all of the
figures that follow. In our previous
studies~\cite{Zenkevich:2018fzl,Zenkevich:2019ayk,Zenkevich:2020ufs,Zenkevich:2022dju}
the ``picture'' directions have been a plane $\mathbb{R}_x \times
\mathbb{R}_y$, but in the present short note it is essential that it
is a cylinder $\mathbb{R}_x \tilde{\times} S^1_y$ with $S^1_y$ of
radius $|\ln \Lambda|$. The tilde is meant to indicate that the
cylinder has a nontrivial monodromy (is ``twisted''), i.e.\ when $y$
goes to $\Lambda y $ the coordinate $x$ goes to $\mu x$ (we use
multiplicative notation, i.e.\ all the pictures are drawn in $(|\ln
x|, |\ln y|)$ coordinates\footnote{The coordinates $x$ and $y$ here
  are thought of as complexified by the $U(1)$ Wilson and 't Hooft
  lines around $S^1_{10}$ of the gauge field living on each brane.}).

The 5-branes of Type IIB string theory are labelled by $(p,q)$-charges
with $(0,1)$ being the D5 brane and $(1,0)$ the NS5 brane. General
$(p,q)$-charges can be thought of as arising from bound states of
several D5 and NS5 branes. The directions of the branes in the
``picture'' cylinder are dictated by their $(p,q)$-charges due to
supersymmetry. 5-branes span two out of three $\mathbb{C}$-planes
along which the equivariant action ($\Omega$-background) is turned
on. Hence, for every $(p,q)$ charge there are three types of 5-branes:
spanning $\mathbb{C}_q \times \mathbb{C}_{t^{-1}}$, $\mathbb{C}_q
\times \mathbb{C}_{t/q}$ and $\mathbb{C}_{t^{-1}} \times
\mathbb{C}_{t/q}$. These are labelled by the lower indices and also by
color: violet, red and blue respectively. The representations
corresponding to 5-branes are Fock representations
e.g.~$\mathcal{F}^{(1,0)}_{q,t^{-1}}(u)$~\cite{AFS}, i.e.\ spaces
generated by the action of commuting creation operators $a_{-n}$ ($n
\in \mathbb{N}$) from the vacuum state $|\varnothing,u\rangle$. The
spectral parameter $u \in \mathbb{C}^{\times}$ of the representation
indicates the position of the 5-brane on the picture
cylinder\footnote{The argument of the complex parameter $u$ is
  provided by the Wilson line around $S^1_{10}$ of the $U(1)$ gauge
  field living on the brane.}. The upper indices of $\mathcal{F}$
represent $(p,q)$-charge of the brane while the lower ones indicate
the pair of $\mathbb{C}$-planes it spans. 5-branes of the same color
sitting at the same value of $\tau$ (the coordinate along
$\mathbb{R}_{\tau}$) can have a triple junction, while the
corresponding Fock representations can be joined using an intertwining
operator (the analogue of an invariant tensor for the DIM
algebra). This leads to the refined topological
vertex~\cite{Awata:2005fa} and $(p,q)$-brane webs~\cite{AFS}.

D3 branes span a single $\mathbb{C}$-plane and therefore are labelled
by a single index and the corresponding color $q$ (blue), $t^{-1}$
(red) or $t/q$ (violet). Classically a D3 brane spans
$\mathbb{R}_{\tau}$ and lives at a point in the picture
cylinder. However, with equivariant deformation turned on the
coordinates of the D3 brane in the picture space become
non-commutative. Hence, it does not have a well-defined position
$(x,y) \in \mathbb{R}_x\times S^1_y$ but instead a wavefunction
depending either on $x$ or on $y$ or some combination
thereof\footnote{In more mathematical terms the wavefunctions should
  be a function on a Lagrangian submanifold of $\mathbb{R}_x\times
  S^1_y$ with respect to the symplectic form $\frac{dx}{x} \wedge
  \frac{dy}{y}$.}. In the current short note we are going to write
everything in the basis of D3 brane wavefunctions with definite $x$
coordinate and draw the D3 branes as vertical dashed lines at fixed
values of $x$. D3 branes do not have $(p,q)$-charges and therefore do
not change the direction of 5-branes upon joining them. The
corresponding DIM representations are called vector representations
$\mathcal{V}$ and they can be thought of as representations on the
space of functions of $x$. We denote the basis states of the
representation $\mathcal{V}$ by $|w\rangle = \delta \left( \frac{x}{w}
\right)$. In fact the only states appearing in $\mathcal{V}_q$ are of
the form $|q^n w\rangle$ with $n \in \mathbb{Z}$.

Branes can join together or pass behind each other in the plane of the
picture. Correspondingly DIM algebra representations can be glued
using an intertwiner or can be exchanged using an $R$-matrix. Many of
these possibilities were explored in~\cite{
  Zenkevich:2018fzl,Zenkevich:2019ayk,AFS,
  Zenkevich:2020ufs,Zenkevich:2022dju}. In the present short note we
clarify how the dictionary between branes and representations works
when the picture space becomes a cylinder. All local objects like
brane junctions and crossings remain the same but the global structure
is different which, as we will see, leads to interesting effects and
novel expressions for the intertwiners.

Since the picture space is a cylinder, the branes can wrap around
it. Here we need to be careful because of the twist of the
cylinder. For example in general it is not possible to wrap an
isolated D5 brane around $S^1_y$ since its $x$ coordinate will change
to $\mu x$ after passing around the circle. If $\mu$ is nontrivial one
needs extra branes with which the D5 would interact in order to glue
the two ends of it. An example of such a picture is given below. Wavy
lines henceforth indicate the identification of the branes passing
through them --- notice that pairs of corresponding wavy lines are
shifted with respect to each other.
\begin{equation}
  \label{eq:55}
   \includegraphics[valign=c]{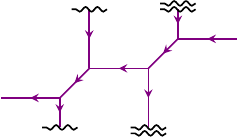}
\end{equation}
In Eq.~(\ref{eq:55}) the ends of the two D5 branes are shifted because
of the interaction with an NS5 brane. The gauge theory engineered by
this setup is the $5d$ $\mathcal{N}=1$ $U(2)$ theory with an extra
adjoint field.

In DIM representation theory compactification of the $y$ direction is
normally interpreted as taking the \emph{trace} over the
representations corresponding to branes wrapping the circle. For
example, in Eq.~(\ref{eq:55}) one needs to take a trace over
$\mathcal{F}^{(0,1)}_{q,t^{-1}}(w_1) \otimes
\mathcal{F}^{(0,1)}_{q,t^{-1}}(w_2)$. The twist $\mu$ and the radius
of the (complexified exponentiated) circumference of the circle
$\Lambda$ are incorporated as the fugacities corresponding to the two
gradings $d$ and $d_{\perp}$ on the DIM algebra (again we invite the
reader to consult the Appendices of~\cite{Zenkevich:2018fzl}
and~\cite{Zenkevich:2020ufs} for notations), so that the relevant
intertwiner has the form
\begin{equation}
  \label{eq:56}
  \Tr_{\mathcal{F}^{(0,1)}_{q,t^{-1}}(w_1) \otimes
\mathcal{F}^{(0,1)}_{q,t^{-1}}(w_2)} \Lambda^d \mu^{d_{\perp}} \left( \ldots \right),
\end{equation}
where $(\ldots)$ denote the interactions with the NS5 brane.

The crucial observation is that there are more general brane
configurations on the cylinder satisfying the conditions of
supersymmetry and corresponding to interesting operators in DIM
representation theory. Those appear when a vertical brane goes full
circle around $S^1_y$ but \emph{does not} return to the same $x$
position. Then the brane can continue to the next circle and so on,
possibly for an infinite number of wrappings. We call such
configurations spiralling branes. Algebraically these are \emph{not
  traces} of the combinations of intertwiners ``along the
compactification direction'', at least not traces of any finite number
of them. Nevertheless the resulting expressions are covariant with
respect to DIM transformations, and this fact ensures their nice
properties.

In this short letter we provide two examples of configurations
involving spiralling branes.

\begin{figure}[h]
  \centering

      \includegraphics[valign=c]{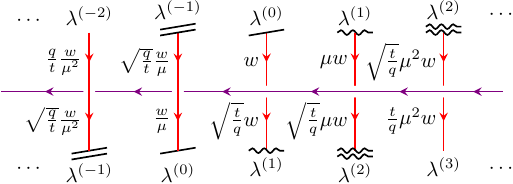}
  
      \caption{The configuration with spiralling D5$_{q,t/q}$ brane
        crossing NS5$_{q,t^{-1}}$ brane. Every turn of the spiral is
        shifted by $\mu$ while the circumference of the circle is
        $\Lambda$. Pairs of wavy or slanted lines indicate
        identification of the branes sticking into them.  Notice that
        the crossings to the left of the center are over, while those
        to the right are under. The Young diagrams $\lambda^{(i)}$ are
        interlacing as in Eq.~\eqref{eq:70}.}
  \label{5-spiral}
\end{figure}

\begin{figure}[h]
    \centering

    \includegraphics[valign=c]{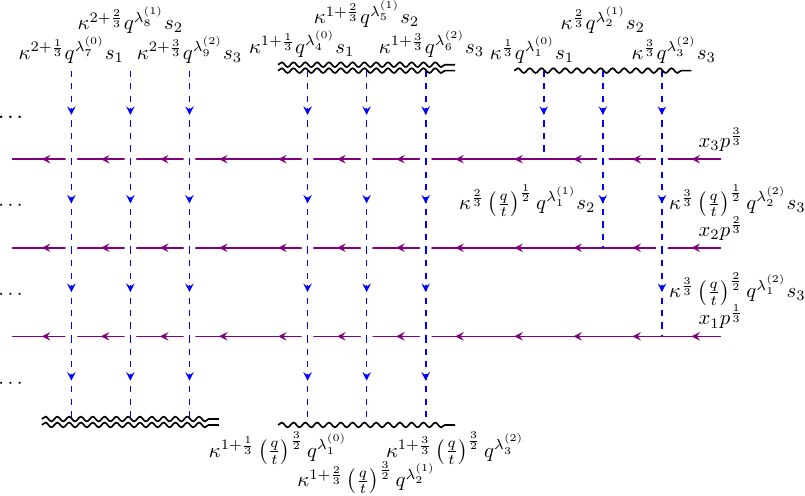}
    
    \caption{The configuration of $N$ spiralling D3$_q$ branes and $N$
      horizontal NS5$_{q,t^{-1}}$ branes ($N=3$ in the picture). In
      order not to clutter the figure not all spectral parameters of
      the branes are shown.}
    \label{fig:3-spiral}
  \end{figure}

\begin{enumerate}
\item The first setup is a D5$_{q,t/q}$ brane spiralling around
  $S^1_y$ \emph{crossing} an NS5$_{q,t^{-1}}$ brane (see
  Fig.~\ref{5-spiral}).  We will find that the matrix element of the
  intertwining operator corresponding to Fig.~\ref{5-spiral} is given
  by the $K$-theoretic vertex function~\cite{Nekrasov:2014nea,
    Okounkov:2015spn} counting sheaves on $\mathbb{C}^3$.

\item The second setup is a stack of $N$ D3$_q$ brane spiralling and
  eventually joining a stack of $N$ NS5$_{q,t^{-1}}$ branes as shown
  in Fig.~\ref{fig:3-spiral}. The resulting operator turns out to
  reproduce the screened vertex operator introduced by Shiraishi
  in~\cite{Shir-1,Langmann:2020utd}. Let us note that for
  \emph{specific} choice of deformation parameters the screened vertex
  operator has been obtained in~\cite{Fukuda:2020czf}. This choice
  corresponds to the case when the D3 brane spiral becomes tighter and
  tighter essentially collapsing into a circle (the situation is
  slightly more complicated, as we discuss in
  sec.~\ref{sec:symm-shiar-wavef}, but the main idea is true).
  
\end{enumerate}
These two examples are meant to demonstrate the usefulness of the
novel concept of spiralling branes. It is already clear that many more
applications can be explored. Also the physical meaning of the
spiralling does not seem to be completely understood; this will be
explored elsewhere.

The rest of the paper is organized as follows. In
sec.~\ref{sec:remind-hanany-witt} we remind some of the results
of~\cite{Zenkevich:2022dju}, in sec.~\ref{sec:spiralling-5-branes} we
compute the matrix elements of the setup from Fig.~\ref{5-spiral} and
show that it coincides with the $K$-theoretic vertex function. In
sec.~\ref{sec:shir-wavef-from} we use the $R$-matrices in the tensor
product of Fock and vector representation to construct the screened
vertex operator which reproduces that of Shiraishi.  We list some
of the ideas for the future in sec.~\ref{sec:conclusions-1}.

  \section{Reminder: Hanany-Witten 5-brane crossing operators}
\label{sec:remind-hanany-witt}
Let us remind the main result of~\cite{Zenkevich:2022dju} (we use the
same notation throughout the paper). It turns out that the $R$-matrix
acting in the tensor product $\mathcal{F}^{(0,1)}_{q,t/q}(w) \otimes
\mathcal{F}^{(1,0)}_{q,t^{-1}}(u)$ is exactly computable and has
interesting combinatorial properties. We call the resulting operator
the Hanany-Witten (HW) brane crossing operator. The $(\lambda,\mu)$
matrix element of the $R$-matrix taken along the vertical Fock
representation in the standard Macdonald polynomial basis is given by:
\begin{multline}
  \label{eq:57}
  \widetilde{\mathcal{R}}^{\lambda}_{\mu}(w,u) = \left( \ldots \otimes
    \left\langle \mu, \sqrt{\frac{t}{q}} w \right| \right)
  \mathcal{R}|_{\mathcal{F}^{(0,1)}_{q,t/q}(w) \otimes
    \mathcal{F}^{(1,0)}_{q,t^{-1}}(u)} \left( | \lambda, w\rangle
    \otimes \ldots \right) =\\
  = \quad \includegraphics[valign=c]{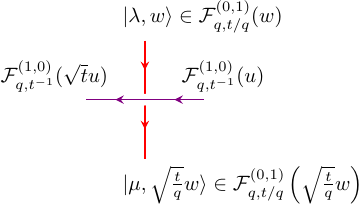}
  \hspace{-3em} =\\
  =\quad u^{|\lambda|-|\mu|} f^{\lambda}_{\mu}
  t^{\frac{\hat{Q}}{2}} \exp \left[ -\sum_{n \geq 1} \frac{1-t^{-n}}{n} \left[ \frac{1}{1-q^n}  + \left( \frac{t}{q} \right)^n \mathrm{Ch}_{\mu}
    \left( q^n, \left( t/q \right)^n \right) -
    \mathrm{Ch}_{\lambda} \left( q^n, \left( t/q \right)^n
    \right) \right] w^n a_{-n}
  \right]\times \\
  \times \exp \left[ \sum_{n \geq 1} \frac{1-t^n}{n} \left[
      \mathrm{Ch}_{\mu} \left( q^{-n}, \left( t/q \right)^{-n} \right)
      - \mathrm{Ch}_{\lambda} \left( q^{-n}, \left( t/q \right)^{-n}
      \right) \right] w^{-n} a_n \right],
\end{multline}
where
\begin{equation}
  \label{eq:58}
  \mathrm{Ch}_{\lambda}(q_1,q_2) = \sum_{(i,j)\in \lambda} q_1^{j-1}
  q_2^{i-1},
\end{equation}
the generators $a_n$ acting on the horizontal Fock space
$\mathcal{F}^{(1,0)}_{q,t^{-1}}(u)$ satisfy
\begin{equation}
  \label{eq:61}
[a_n, a_m] = \frac{1 - q^{|n|}}{1 - t^{|n|}} n \delta_{n+m,0},
\end{equation}
and the prefactor $f^{\lambda}_{\mu}$ is nonvanishing only when the
partitions $\lambda$ and $\mu$ are \emph{interlacing:}
\begin{equation}
  \label{eq:59}
  f^{\lambda}_{\mu} = 0 \quad \text{for } \lambda \nsucc \mu
\end{equation}
which means
\begin{equation}
  \label{eq:62}
  \lambda \succ \mu \qquad \Leftrightarrow \qquad  \ldots \geq \lambda_i \geq \mu_i \geq \lambda_{i+1} \geq \mu_{i+1}
  \geq \ldots
\end{equation}
The operator $t^{\hat{Q}/2}$ shifts the spectral parameter $u$ of the
horizontal Fock space by $\sqrt{t}$:
\begin{equation}
  \label{eq:60}
    t^{\hat{Q}/2} | u, \alpha \rangle = | \sqrt{t} u, \alpha \rangle. 
\end{equation}
The interlacing condition~\eqref{eq:62} endows the
$R$-matrix~\eqref{eq:57} with nice combinatorial properties. We will
use them when we combine several brane crossing operators together in
a chain.

In addition to the formula~\eqref{eq:57} which can be called the
undercrossing (as evident from the picture) we can derive the formula
for \emph{overcrossing,} i.e.\ when the branes pass each other in
different order in the $\tau$ coordinate. This corresponds to
evaluating the inverse of the $R$-matrix. We find
\begin{multline}
   \label{eq:63}
 (\widetilde{\mathcal{R}}^{-1})^{\lambda}_{\mu}(w,u) = \left( \ldots \otimes
    \left\langle \mu, \sqrt{\frac{q}{t}} w \right| \right)
  \mathcal{R}^{-1}|_{\mathcal{F}^{(0,1)}_{q,t/q}(w) \otimes
    \mathcal{F}^{(1,0)}_{q,t^{-1}}(u)} \left( | \lambda, w\rangle
    \otimes \ldots \right) =\\
  = \quad \includegraphics[valign=c]{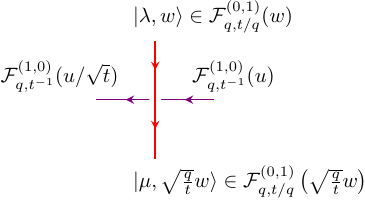}
  \hspace{-3em} =\\
  =\quad u^{|\lambda|-|\mu|} g^{\lambda}_{\mu}
  t^{-\frac{\hat{Q}}{2}} \exp \left[ -\sum_{n \geq 1} \frac{1-t^{-n}}{n} \left[ \mathrm{Ch}_{\mu}
    \left( q^n, \left( t/q \right)^n \right) -
    \mathrm{Ch}_{\lambda} \left( q^n, \left( t/q \right)^n
    \right) \right] w^n a_{-n}
  \right]\times \\
  \times \exp \left[ \sum_{n \geq 1} \frac{1-t^n}{n} \left(
      \frac{t}{q} \right)^n \left[ - \frac{1}{1-q^{-n}} +
      \mathrm{Ch}_{\mu} \left( q^{-n}, \left( t/q \right)^{-n} \right)
      -  \left( \frac{q}{t} \right)^n \mathrm{Ch}_{\lambda} \left( q^{-n}, \left( t/q \right)^{-n}
      \right) \right] w^{-n} a_n \right],
\end{multline}
where the prefactor $g^{\lambda}_{\mu}$ now requires $\mu$ to
interlace $\lambda$, not vice versa:
\begin{equation}
  \label{eq:64}
    g^{\lambda}_{\mu} = 0 \quad \text{for } \mu \nsucc \lambda.
\end{equation}
Notice that the expression~\eqref{eq:63} is a generalization of
Eq.~(67) from~\cite{Zenkevich:2020ufs} to which it reduces for
$\lambda = \mu = \varnothing$.

One can also write down the Hanany-Witten (HW)
\emph{move}~\cite{Hanany:1996ie} relating the over- and undercrossing
operators~\eqref{eq:57} and~\eqref{eq:63}. The key combinatorial
observation is that if $\lambda \succ \mu$, then $\mu \succ [\lambda
\backslash \lambda_1]$, where $[\lambda \backslash \lambda_1]$ denotes
the Young diagram obtained from $\lambda$ by removing the first
column. We then have
\begin{equation}
  \label{eq:65}
  (\widetilde{\mathcal{R}})^{\lambda}_{\mu}(w,u) \sim
  \Phi^q_{q,t^{-1}}(q^{\lambda_1} w) (\widetilde{\mathcal{R}}^{-1})^{[\lambda \backslash
    \lambda_1]}_{\mu}\left(\frac{t}{q} w,u\right), 
\end{equation}
where $\Phi^q_{q,t^{-1}}(w)$ is the D3$_q$-NS5$_{q,t^{-1}}$ brane
junction operator~\cite{Zenkevich:2018fzl}:
\begin{equation}
  \label{eq:66}
  \Phi^q_{q,t^{-1}}(q^{\lambda_1} w) \sim u^{\lambda_1} t^{\hat{Q}}
  :\exp \left[ \sum_{n \neq 0} \frac{1-t^n}{n}
    \frac{q^{-n\lambda_1}}{1-q^{-n}} a_n w^{-n} \right]:.
\end{equation}
Graphically one can picture the HW move as follows:
\begin{equation}
  \label{eq:67}
 \includegraphics[valign=c]{figures/d5-ns5-hw-lm-crop} \quad
  = \quad   \includegraphics[valign=c]{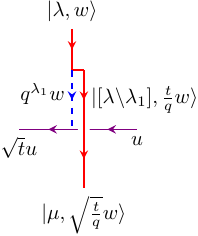}
\end{equation}
The junction between the dashed blue line and the vertical red line is
a trivial operator cutting the Young diagram $\lambda$ into the first
column $\lambda_1$ and the rest of the diagram $[\lambda \backslash
\lambda_1]$.

As a side remark let us notice a curious fact that the Hanany-Witten
\emph{rule}~\cite{Hanany:1996ie} also holds for the crossing and
junction operators that we have introduced. The rule states that
between a given pair of D5$_{q,t/q}$ and NS5$_{q,t^{-1}}$ branes there
can stretch at most one D3$_q$ brane. Indeed, one can evaluate the
operator corresponding to the following picture
\begin{equation}
  \label{eq:68}
  \includegraphics[valign=c]{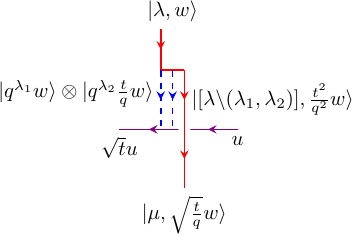} \quad \sim
  \quad \frac{\left( q^{\lambda_2 - \lambda_1} ;q \right)_{\infty}}{\left(t q^{\lambda_2 - \lambda_1} ;q \right)_{\infty}}
\end{equation}
and learn that it vanishes since $\lambda_1 \geq \lambda_2$ in a Young
diagram.

\section{Spiralling 5-branes and the $K$-theoretic vertex}
\label{sec:spiralling-5-branes}
As noted in~\cite{Zenkevich:2022dju} the interlacing
conditions~\eqref{eq:59}, \eqref{eq:64} for the Young diagrams on the
legs of the HW crossings~\eqref{eq:57}, \eqref{eq:63} remind one of
diagonal slices of a $3d$ Young diagram (plane partitions). Indeed, if
$\pi = \{\pi^{(i)}_j\}$, $i,j \in \mathbb{N}$ is a plane partition, so
that $\pi^{(i)}_j \geq \pi^{(i+1)}_j$, $\pi^{(i)}_j \geq
  \pi^{(i)}_{j+1}$, then the set of Young diagrams $\{
    \lambda^{(i)}\}$ defined by
\begin{equation}
   \label{eq:69}
  \lambda^{(i)}_j = \pi^{(j)}_{i+j-1}, \quad i \in \mathbb{Z}, j
  \in \mathbb{N}.  
\end{equation}
satisfies the interlacing conditions
\begin{equation}
  \label{eq:70}
\ldots \prec \lambda^{(-3)}  \prec \lambda^{(-2)}  \prec \lambda^{(-1)}  \prec \lambda^{(0)} \succ \lambda^{(1)} \succ
  \lambda^{(2)} \succ \lambda^{(3)} \succ \ldots 
\end{equation}
Notice that the infinite tails in the sequence~\eqref{eq:70} may have
nontrivial asymptotics.

This makes one wonder if the sequence of Young diagrams~\eqref{eq:70}
can be understood as living on the legs of a chain of HW crossing
operators. It is not hard to guess the right configuration of branes,
which turns out to be a spiralling D5$_{q,t/q}$ brane crossing a
horizontal NS5$_{q,t^{-1}}$ brane, as shown in
Fig.~\ref{5-spiral}. The Figure depicts an intertwining of the DIM
algebra involving an infinite chain of $R$-matrices in which the sum
over intermediate states on the vertical Fock representation
corresponds to the sum over \emph{plane partitions.} Let us emphasize this is \emph{not} the sum over plane partitions
featuring in the (refined) topological vertex/triple 5-brane junction,
each slice here originates from its own segment of the D5$_{q,t/q}$
brane and it is essential that the number of segments (and crossings)
is infinite.

The algebraic expression for the intertwiner in Fig.~\ref{5-spiral} is
\begin{multline}
  \label{eq:74}
\mathfrak{R} = \left[\prod_{i=-\infty}^{-1}  (\Lambda^{id}
    \mu^{i d^{\perp}} \otimes 1) \mathcal{R}^{-1} (\Lambda^{(1-i)d}
    \mu^{(1-i) d^{\perp}} \otimes 1) \right] \mathcal{R}\times\\
  \times\left. \left[ \prod_{i=1}^{\infty}
   (\Lambda^{id} \mu^{i d^{\perp}} \otimes 1) \mathcal{R}(\Lambda^{(1-i)d} \mu^{(1-i) d^{\perp}} \otimes 1) \right]
\right|_{\mathcal{F}^{(0,1)}_{q,t/q} \otimes \mathcal{F}^{(1,0)}_{q,t^{-1}}}.
\end{multline}
In fact $\mathfrak{R}$ is a \emph{Drinfeld twist} of the standard DIM
$R$-matrix using the cocycle $F= \left[ \prod_{i=1}^{\infty}
  (\Lambda^d \mu^{d^{\perp}} \otimes 1) \mathcal{R} \right]$. Since
$\mathfrak{R}$ acts on the tensor product of two Fock spaces its
matrix elements depend on four Young diagrams --- the asymptotics of
the sequences of states on the vertical and horizontal branes.

The next logical question is what natural object can be associated
with the sum over plane partitions? A very interesting candidate is
the $K$-theoretic vertex, the equivariant count of sheaves on
$\mathbb{C}^3$~\cite{Nekrasov:2014nea,Okounkov:2015spn}. The vertex is
given by an explicit formula:
\begin{equation}
  \label{eq:71}
  I(\lambda,\mu,\nu) = \sum_{\pi} Q^{|\pi|'} \exp \left[ \sum_{n \geq 1}
    \frac{1}{n} \chi_{\pi}(t_1^n, t_2^n, t_3^n) \right],
\end{equation}
where the sum goes over plane partitions with fixed asymptotics
$(\lambda,\mu,\nu)$ along three coordinate axes, $|\pi|'$ denotes the
regularized total number of boxes in $\pi$,
\begin{equation}
  \label{eq:72}
  \chi_{\pi}(t_1, t_2, t_3) = \mathrm{Ch}_{\pi}(t_1, t_2, t_3) -
  \frac{\mathrm{Ch}_{\pi}(t_1^{-1}, t_2^{-1},
  t_3^{-1})}{t_1 t_2 t_3}  + \mathrm{Ch}_{\pi}(t_1, t_2, t_3) \mathrm{Ch}_{\pi}(t_1^{-1}, t_2^{-1}, t_3^{-1})
  \frac{(1-t_1)(1-t_3)(1-t_3)}{t_1 t_2 t_3} 
\end{equation}
and
\begin{equation}
  \label{eq:73}
  \mathrm{Ch}_{\pi}(t_1, t_2 ,t_3) = \sum_{(i,j,k)\in \pi} t_1^{j-1}
  t_2^{i-1} t_3^{k-1}.
\end{equation}
We find quite remarkable that the following theorem holds.

\begin{thm}
  Up to $\Lambda$- and $\mu$-independent prefactors the $K$-theoretic
  vertex with two nontrivial asymptotics coincides with the
  vacuum matrix element of $\mathfrak{R}$
  \begin{equation}
    \label{eq:75}
    (\langle \varnothing | \otimes \langle\mu| ) \mathfrak{R} (|\lambda\rangle
    \otimes | \varnothing \rangle) \sim I(\lambda,\mu,\varnothing)
  \end{equation}
  provided one identifies
  \begin{align}
    \label{eq:76}
    t_1 &= \mu^{-1} \sqrt{\frac{t}{q}},\\
    t_2 &= \mu \sqrt{\frac{t}{q}},\\
    t_3 &= q,\\
    Q &= \Lambda.
  \end{align}
\end{thm}
The proof is straightforward term-by-term analysis of the matrix
elements of $\mathfrak{R}$. Several remarks are in order.
\begin{enumerate}
\item Notice that $t_1 t_2 t_3 = t$, so one can nicely write three
  independent parameters $(t_1,_2,t_3)$ in terms of four parameters
  $(t_1, t_2 ,t_3, (t_1 t_2 t_3)^{-1}) = \left( \mu^{-1}
    \sqrt{\frac{t}{q}}, \mu \sqrt{\frac{t}{q}} , q, t^{-1} \right)$
  with product equal to one.

\item In the definition of $\mathfrak{R}$ the built-in symmetry of the
  DIM algebra under permutations of $(q,t^{-1},t/q)$ is
  broken. Instead the new symmetry between $(t_1, t_2, t_3)$ emerges
  in a completely nontrivial way.

\item It is not hard to add the third nontrivial leg to the vertex by
  considering instead of $\mathfrak{R}$, i.e.\ a product of the form $
  \cdots \mathcal{R}^{-1} \mathcal{R}^{-1} \mathcal{R}^{-1}
  \mathcal{R} \mathcal{R} \mathcal{R} \cdots$, a more general product
  of $\pm 1$ powers with the same asymptotics. Such a pattern of $\pm
  1$ powers encodes a Maya diagram which is equivalent to a Young
  diagram, which is the diagram living on the third leg in the
  $K$-theoretic vertex.

\item The infinite wrapping of the D5$_{q,t/q}$ brane and the
  appearance of $\mu$ as an extra equivariant parameter in the vertex
  suggest that the extra sums over an infinite number of Young
  diagrams could be rewritten as sums over Kaluza-Klein modes in the
  $S^1_y$ direction. It would be interesting to make this point more
  explicit in the future.

\item Non-vacuum matrix elements of $\mathfrak{R}$ should give
  $K$-theoretic vertex with descendant insertions.
\end{enumerate}

\section{Shiraishi wavefunctions from spiralling D3-branes}
\label{sec:shir-wavef-from}
In~\cite{Shir-1,Langmann:2020utd} Shiraishi has introduced a new class
of special functions originally called ``non-stationary elliptic
Ruijsenaars wavefunctions''. We will call them simply Shiraishi
functions for brevity. These functions can be obtained as matrix
elements of a certain screened vertex operator involving affine system
of screening currents. In this section we show that the affine
screening operators featuring in the Shiraishi vertex operator
naturally arise in the wrapped background with spiralling D3$_q$
branes.

Consider a tensor product of $N$ horizontal Fock spaces
$\mathcal{F}^{(1,0)}_{q,t^{-1}}\left(p^{\frac{i}{N}} x_i\right)
$. Then according to the general formalism~\cite{Zenkevich:2018fzl}
there are screening currents between neighbouring pairs of Fock spaces
built from intertwiners of DIM algebra involving vector
representations $\mathcal{V}_q$ or $\mathcal{V}_{t^{-1}}$ (these are
usually called dual sets of screening charges). However, the essential
difference from~\cite{Zenkevich:2018fzl} arises because the space is
now compactified, so there appears an extra screening charge
corresponding to the interaction between the first and the last Fock
spaces in the tensor product.

The Shiraishi wavefunction $f^{\widehat{\mathfrak{gl}}_N}$ depends on
two sets of $N$ variables $\mathbf{x} = \{x_i\}$ and $\mathbf{s} =
\{s_i\}$ and parameters $p$, $\kappa$, $q$ and $t$. It can be defined
as a vacuum matrix element of a vertex operator as follows:
\begin{equation}
  \label{eq:77}
f^{\widehat{\mathfrak{gl}}_N} (\mathbf{x}, p^{\frac{1}{N}}| \mathbf{s},
  \kappa^{\frac{1}{N}}|q,t) = \prod_{1 \leq i < j \leq N} \frac{\left(
      \kappa^{\frac{j-i}{N}} t  \frac{s_j}{s_i} ;q,\kappa
    \right)_{\infty} \left(
      \kappa^{\frac{j-i}{N}} \frac{q}{t} \frac{s_j}{s_i} ;q,\kappa \right)_{\infty}}{\left(
      \kappa^{\frac{j-i}{N}} \frac{s_j}{s_i} ;q,\kappa
    \right)_{\infty} \left(
      \kappa^{\frac{j-i}{N}} q \frac{s_j}{s_i} ;q,\kappa \right)_{\infty}} \langle \varnothing |
  \prod_{i=0}^{N-1} \Phi^i(s_{i+1}|t^{\omega^{N-i}} \mathbf{x}, p) |
    \varnothing\rangle ,
  \end{equation}
where
\begin{align}
  \label{eq:78}
  (x;q,p)_{\infty} &= \prod_{i,j \geq 0} (1 - x q^i p^j),\\
  t^{\omega^m} \mathbf{x} &= (x_1, x_2 , \ldots, t x_{N-i+1}, \ldots ,
    t x_N )\\
    \Phi^i(s|t^{\omega^{N-i}}\mathbf{x},p) &= \sum_{\lambda}
    \Phi_{\lambda}^i (s) \left( p^{\frac{1}{N}} \prod_{j \geq 1}
      \frac{t^{\delta_{j \mod N,1}} x_{(N-i+j) \mod N}}{x_{(N-i + j-1)
          \mod N}}
    \right)^{\lambda_j},\\
    \Phi^i_{\lambda}(s) &= \left(\frac{\left( \frac{q}{t};q
        \right)_{\infty}}{\left( q;q
        \right)_{\infty}}\right)^{l(\lambda)} \phi_{i-l(\lambda)}(s)
    \prod_{j=1}^{l(\lambda)} S_{j-i+1}\left(
        \kappa^{\frac{l(\lambda)}{N}} q^{\lambda_j} s
        \right),\\
        \phi_i(s) &= :\exp \left[ \sum_{n \neq 0} \frac{w^{-n}}{n}
          \beta_n^{(i)}\right]:,\label{eq:3}\\
        S_i(w) &= :\exp \left[ -\sum_{n \neq 0} \frac{w^{-n}}{n}
          \alpha^{(i)}_n \right]:,\label{eq:2}
      \end{align}
      and the bosonic modes $\alpha^{(i)}_n$, $\beta^{(i)}_n$ satisfy
      certain Heisenberg commutation relations based on the affine
      root system. The exact expressions for these commutation
      relations can be found in~\cite{Shir-1}.

      The main theorem in this section states that one can reproduce
      all the screening currents and vertex operators in~\eqref{eq:3}
      and~\eqref{eq:2} if one considers the system of spiralling
      branes depicted in Fig.~\ref{fig:3-spiral}.
      
\begin{thm}
  \label{thm:shir-wavef-from-1}
  The vacuum matrix element in the r.h.s.~of~\eqref{eq:77} coincides
  with the vacuum matrix element of the intertwiner shown in
  Fig.~\ref{fig:3-spiral}. The map between the parameters is already
  incorporated in the Figure.
\end{thm}
The proof is tedious but straightforward and involves checking the
commutation relations of the screening currents and establishing the
map between the parameters of the Shiraishi functions and the spectral
parameters of the representations in the DIM intertwiner. In fact,
more can be said: not only does the vacuum matrix elements coincide,
but the screened vertex operator itself is the same as the intertwiner
from Fig.~\ref{fig:3-spiral}.

Notice that Theorem~\ref{thm:shir-wavef-from-1} provides the
identification between the network of DIM intertwiners and the
wavefunction for completely general parameters. One can then follow
the dictionary between the parameters for any possible degeneration,
such as the one considered in~\cite{Fukuda:2020czf}.

Having new algebraic view on Shiraishi functions gives us hints about
their properties. In the next section we provide as just one example,
the sketch of a proof of the mirror symmetry of the functions
$f^{\widehat{\mathfrak{gl}}_N}$.

\section{Symmetries of Shiarishi wavefunctions from spiralling branes}
\label{sec:symm-shiar-wavef}
Functions $f^{\widehat{\mathfrak{gl}}_N} (\mathbf{x}, p^{\frac{1}{N}}|
\mathbf{s}, \kappa^{\frac{1}{N}}|q,t)$ have nontrivial symmetry
properties which are not evident from their definition. In partocular,
they enjoy the so-called mirror symmetry (the exact prefactor is also
known~\cite{Shir-1}, but we will not pursue it here):
\begin{equation}
  \label{eq:4}
  f^{\widehat{\mathfrak{gl}}_N} (\mathbf{x}, p^{\frac{1}{N}}|
\mathbf{s}, \kappa^{\frac{1}{N}}|q,t) \sim   f^{\widehat{\mathfrak{gl}}_N} (\mathbf{s}, \kappa^{\frac{1}{N}}|\mathbf{x}, p^{\frac{1}{N}}
|q,q/t).
\end{equation}

In the spiralling branes formalism the symmetry~\eqref{eq:4} becomes
the consequence of the HW move~\eqref{eq:67}. Indeed, after a little
manipulation (attaching the free ends of the D3$_q$ branes to
D5$_{q,t/q}$ branes, which adds only a prefactor to the partition
function~\cite{Zenkevich:2020ufs}) one learns that
Fig.~\ref{fig:3-spiral} is equivalent to the following picture:
\begin{equation}
  \label{eq:5}
  \includegraphics[valign=c]{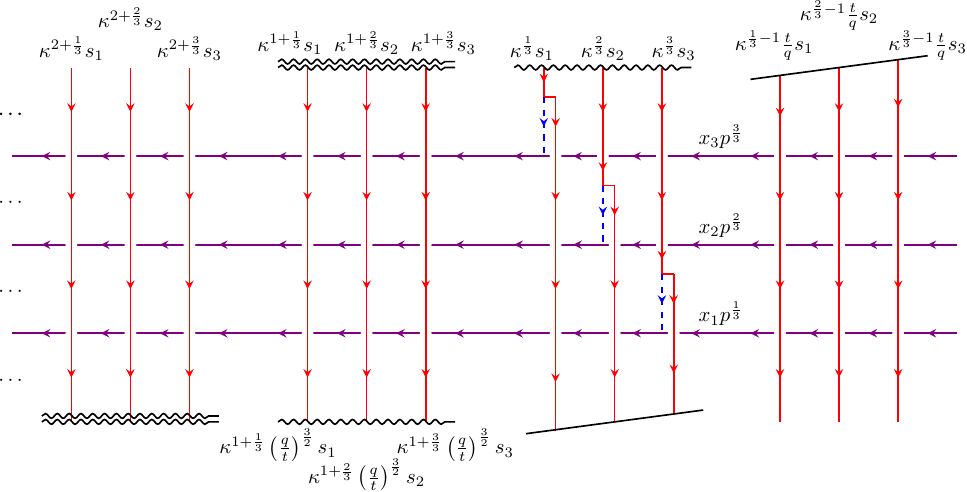} 
\end{equation}
Using the HW move~\eqref{eq:67} we can eliminate D3$_q$ branes
altogether and encode their positions in the pattern of over- and undercrossings:
\begin{equation}
  \label{eq:6}
  \includegraphics[valign=c]{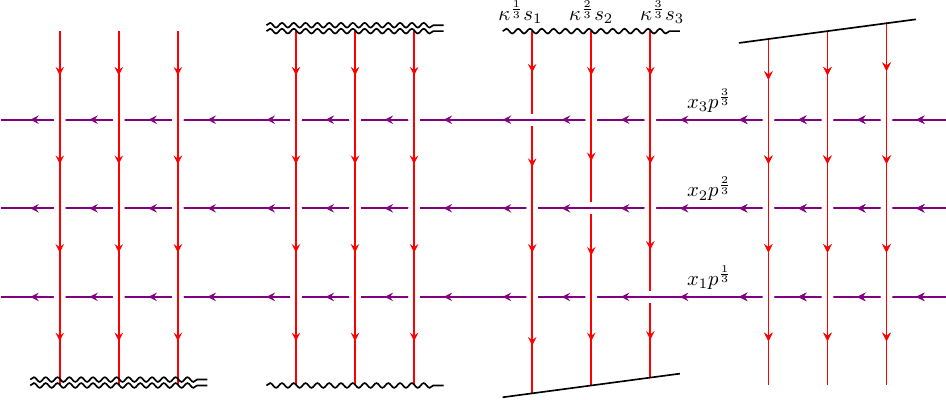} 
\end{equation}

It might not be evident from the first sight, but after a little
redrawing one can see that the picture~\eqref{eq:6} is symmetric under
the exchange of the vertical and horizontal axes and simultaneous swap
of red and violet colors (i.e.\ $t^{-1}$ and $t/q$ parameters). This
proves the mirror symmetry of Shiraishi wavefunctions.

Spectral duality of the Shiraishi wavefunctions can also be proven
along the same lines, the details of this will appear elsewhere.

\section{Conclusions}
\label{sec:conclusions-1}
We have introduced new types of brane configurations in Type IIB
string theory on the twisted product $\mathbb{R}^8 \tilde{\times}
T^2$. In these configurations a brane wraps one of the cycles of the
torus multiple times while with every wrapping one of the flat
coordinates is shifted by a constant. It turns out that such brane
configurations are meaningful also in the algebraic formalism in which
they correspond to certain DIM algebra intertwiners.

To justify the usefulness of the spiralling branes we considered two
examples in which the corresponding intertwining operators reproduce
beautiful objects ($K$-theoretic vertex function and Shiraishi
wavefunctions) that are both of high importance to general
mathematical physics.

We hope that this approach will lead to many new results, however as
for now we have only scratched the surface. Nevertheless we have
proven mirror symmetry of Shiraishi wavefunctions which turns out to
be natural in our formalism. With the new machinery it would also be
desirable to understand Hamiltonians corresponding to ``nonstationary
elliptic Ruijsenaars'' system.

Let us comment on possible connection with the natural generalization
of the $K$-theoretic vertex function --- the ``magnificent four''
vertex function~\cite{Nekrasov:2020zli}, which is a partition function
counting $4d$ Young diagrams (solid partitions). We conjecture based
on the combinatorics of the crossing operators that the setup similar
to that of Fig.~\ref{5-spiral} but with a D7 brane instead of the D5
brane will in fact reproduce the magnificent four vertex. In the
alebraic language this corresponds to wrapping the MacMahon
representation instead of the Fock one on the spiral. One can also see
that a similar crossing but with D5$_{q,t^{-1}}$ crossing
NS5$_{q,t^{-1}}$ leads to a partition function that can be understood
as a particular reduction of the magnificent four vertex in which the
$4d$ Young diagrams are confined to a ``hook'' in one of the
coordinate planes.

Very recently, $4d$ Young diagrams have also been considered as states
of BPS particles arising from the bound states of
branes~\cite{Galakhov:2023vic}. It would be interesting to incorporate
them into our algebraic description.

An explicit closed form expression for the $K$-theoretic vertex with
two nontrivial legs was obtained in~\cite{Kononov:2019fni}. Our
approach does not seem to give a reason for the existence of such an
expression, however this deserves further investigation.

We also expect that spiralling branes can be used in the new algebraic
approach to $qq$-characters~\cite{Bayindirli:2023byn} to accommodate
gauge theories with adjoint matter.

\section*{Acknowledgements}
\label{sec:acknowledgements}
The author has been supported by the European Research Council under
the European Union’s Horizon~2020 research and innovation programme
under grant agreement No~948885 and by the Royal Society University
Research Fellowship.


\begin{thebibliography}{99}
\bibitem{Polchinski:1998rq}
J.~Polchinski,
Cambridge University Press, 2007,
ISBN 978-0-511-25227-3, 978-0-521-67227-6, 978-0-521-63303-1
doi:10.1017/CBO9780511816079

\bibitem{Sharpe:2003dr}
E.~Sharpe,
[arXiv:hep-th/0307245 [hep-th]].

\bibitem{Harvey:1996gc}
J.~A.~Harvey and G.~W.~Moore,
Commun. Math. Phys. \textbf{197} (1998), 489-519
doi:10.1007/s002200050461
[arXiv:hep-th/9609017 [hep-th]].

  \bibitem{Zenkevich:2018fzl}
Y.~Zenkevich,
JHEP \textbf{08} (2021), 149
doi:10.1007/JHEP08(2021)149
[arXiv:1812.11961 [hep-th]].

\bibitem{DIM} J. Ding, K.
  Iohara, 
  Lett. Math. Phys. {\bf 41} (1997) 181--193, q-alg/9608002\\
  K. Miki, J. Math. Phys. {\bf 48} (2007) 123520

\bibitem{Zenkevich:2019ayk}
Y.~Zenkevich,
JHEP \textbf{12} (2021), 034
doi:10.1007/JHEP12(2021)034
[arXiv:1912.13372 [hep-th]].

\bibitem{Gaiotto:2017euk}
D.~Gaiotto and M.~Rap\v{c}\'ak,
JHEP \textbf{01} (2019), 160
doi:10.1007/JHEP01(2019)160
[arXiv:1703.00982 [hep-th]].\\
T.~Proch\'azka and M.~Rap\v{c}\'ak,
JHEP \textbf{11} (2018), 109
doi:10.1007/JHEP11(2018)109
[arXiv:1711.06888 [hep-th]].\\
T.~Proch\'azka and M.~Rap\v{c}\'ak,
JHEP \textbf{05} (2019), 159
doi:10.1007/JHEP05(2019)159
[arXiv:1808.08837 [hep-th]].\\
M.~Rap\v{c}\'ak,
JHEP \textbf{01} (2020), 042
doi:10.1007/JHEP01(2020)042
[arXiv:1910.00031 [hep-th]].

\bibitem{Galakhov:2018lta}
D.~Galakhov,
Nucl. Phys. B \textbf{946} (2019), 114693
doi:10.1016/j.nuclphysb.2019.114693
[arXiv:1812.05801 [hep-th]].\\
D.~Galakhov and M.~Yamazaki,
Commun. Math. Phys. \textbf{396} (2022) no.2, 713-785
doi:10.1007/s00220-022-04490-y
[arXiv:2008.07006 [hep-th]].\\
D.~Galakhov, W.~Li and M.~Yamazaki,
JHEP \textbf{08} (2021), 146
doi:10.1007/JHEP08(2021)146
[arXiv:2106.01230 [hep-th]].


    \bibitem{Awata:2005fa}
H.~Awata and H.~Kanno,
JHEP \textbf{05} (2005), 039
doi:10.1088/1126-6708/2005/05/039
[arXiv:hep-th/0502061 [hep-th]].\\
  A.~Iqbal, C.~Kozcaz and C.~Vafa,
  JHEP {\bf 0910} (2009) 069 doi:10.1088/1126-6708/2009/10/069
  [hep-th/0701156].

    \bibitem{AFS}
  H.~Awata, B.~Feigin and J.~Shiraishi,
  JHEP {\bf 1203} (2012) 041
  doi:10.1007/JHEP03(2012)041
  [arXiv:1112.6074 [hep-th]].

\bibitem{Zenkevich:2020ufs}
Y.~Zenkevich,
JHEP \textbf{12} (2021), 027
doi:10.1007/JHEP12(2021)027
[arXiv:2012.15563 [hep-th]].

  \bibitem{Zenkevich:2022dju}
Y.~Zenkevich,
[arXiv:2212.14808 [hep-th]].

\bibitem{Nekrasov:2014nea}
N.~Nekrasov and A.~Okounkov,
doi:10.14231/AG-2016-015
[arXiv:1404.2323 [math.AG]].

\bibitem{Okounkov:2015spn}
A.~Okounkov,
[arXiv:1512.07363 [math.AG]].

\bibitem{Shir-1} J.~I.~Shiraishi, 
  Journal of Integrable Systems, 4(1) (2019):
  xyz010.
  [arXiv:1903.07495 [math.QA]]

\bibitem{Langmann:2020utd}
E.~Langmann, M.~Noumi and J.~Shiraishi,
SIGMA \textbf{16} (2020), 105
doi:10.3842/SIGMA.2020.105
[arXiv:2006.07171 [math-ph]].

\bibitem{Fukuda:2020czf}
M.~Fukuda, Y.~Ohkubo and J.~Shiraishi,
SIGMA \textbf{16} (2020), 116
doi:10.3842/SIGMA.2020.116
[arXiv:2002.00243 [math.QA]].

\bibitem{Harada:2021xnm}
K.~Harada, Y.~Matsuo, G.~Noshita and A.~Watanabe,
JHEP \textbf{04} (2021), 202
doi:10.1007/JHEP04(2021)202
[arXiv:2101.03953 [hep-th]].


\bibitem{Hanany:1996ie}
A.~Hanany and E.~Witten,
Nucl. Phys. B \textbf{492} (1997), 152-190
doi:10.1016/S0550-3213(97)00157-0
[arXiv:hep-th/9611230 [hep-th]].


\bibitem{Awata:2016mxc}
H.~Awata, H.~Kanno, A.~Mironov, A.~Morozov, A.~Morozov, Y.~Ohkubo and Y.~Zenkevich,
JHEP \textbf{10} (2016), 047
doi:10.1007/JHEP10(2016)047
[arXiv:1608.05351 [hep-th]].\\
H.~Awata, H.~Kanno, A.~Mironov, A.~Morozov, A.~Morozov, Y.~Ohkubo and Y.~Zenkevich,
Nucl. Phys. B \textbf{918} (2017), 358-385
doi:10.1016/j.nuclphysb.2017.03.003
[arXiv:1611.07304 [hep-th]].\\
H.~Awata, H.~Kanno, A.~Mironov, A.~Morozov, K.~Suetake and Y.~Zenkevich,
JHEP \textbf{04} (2019), 097
doi:10.1007/JHEP04(2019)097
[arXiv:1810.07676 [hep-th]].


\bibitem{Jimbo:1999zz}
M.~Jimbo, H.~Konno, S.~Odake and J.~Shiraishi,
Transform. Groups \textbf{4} (1999), 303-327
doi:10.1007/BF01238562
[arXiv:q-alg/9712029 [math.QA]].


\bibitem{Ghoneim:2020sqi}
M.~Ghoneim, C.~Koz\c{c}az, K.~Kur\c{s}un and Y.~Zenkevich,
Nucl. Phys. B \textbf{978} (2022), 115740
doi:10.1016/j.nuclphysb.2022.115740
[arXiv:2012.15352 [hep-th]].


\bibitem{Bayindirli:2023byn}
M.~B.~Bay\i{}nd\i{}rl\i{}, D.~N.~Demirta\c{s}, C.~Koz\c{c}az and Y.~Zenkevich,
[arXiv:2310.02587 [hep-th]].

\bibitem{Kononov:2019fni}
Y.~Kononov, A.~Okounkov and A.~Osinenko,
Commun. Math. Phys. \textbf{382} (2021) no.3, 1579-1599
doi:10.1007/s00220-021-03936-z
[arXiv:1905.01523 [math-ph]].

\bibitem{Nekrasov:2020zli}
N.~Nekrasov,
Ann. Inst. H. Poincare D Comb. Phys. Interact. \textbf{7} (2020) no.4, 505-534
doi:10.4171/aihpd/93\\
N.~Nekrasov and N.~Piazzalunga,
[arXiv:2306.12995 [hep-th]].

\bibitem{Galakhov:2023vic}
D.~Galakhov and W.~Li,
[arXiv:2311.02751 [hep-th]].



\end{thebibliography}
\end{document}